%% file: main.tex
\PassOptionsToPackage{colorlinks=true, citecolor=Black, urlcolor=Black, linkcolor=Black}{hyperref}
\documentclass[journal=acsau, manuscript=article, doi=true, layout=traditional]{achemso}
\usepackage{amsmath, amssymb, array, booktabs, hypdoc, listings, lmodern, mathpazo, microtype}
\usepackage[dvipsnames]{xcolor}
\usepackage{sidecap}
\bibstyle{achemso}

\author{Abhishek Grewal}
\email{a.grewal@fkf.mpg.de}
\author{Christopher C. Leon}
\email{ccleon@mit.edu}
\author{Olle Gunnarsson}
\email{o.gunnarsson@fkf.mpg.de}
\affiliation{Max-Planck-Institut f{\"u}r Festk{\"o}rperforschung, Heisenbergstra{\ss}e 1, 70569 Stuttgart, Germany}

\title{Scanning Tunneling Microscopy for Molecules: Manipulating Electron Transport through the Conduction Gap by varying Buffer Layer}
\date{\today}


\begin{document}

\begin{abstract}
\noindent In scanning tunneling microscopy of molecules, an insulating buffer layer is often introduced to reduce interactions between adsorbed molecules and the substrate.
Focusing on tunneling through the molecule's electronic transport gap, we demonstrate that the buffer itself strongly influences the wave function of the tunneling electron at the molecule.
This is exemplified for an adsorbed platinum phthalocyanine molecule by varying the composition and thickness of the buffer.
We find that, in particular, the buffer's lattice parameter is crucial.
By expanding the wave function of the tunneling electron in molecular orbitals (MOs), we illustrate how one can strongly vary the relative weights of different MOs, such as the highest occupied MO versus some low-lying MOs with few nodal surfaces.
The set of MOs with significant weight are important for processes used to manipulate the state of the molecule by a tunneling electron, such as molecular luminescence. The choice of buffer therefore provides an important tool for manipulating these processes.
\end{abstract}

\bigskip
%\section{Introduction}\label{sec:1}
\noindent
Scanning tunneling microscopy (STM) has been extensively used to study the intrinsic electronic structure of adsorbed molecules \cite{binnigTunnelingControllable1982, binnigSurfaceStudies1982, binnigIfmmodeTimes1983, binnig111Facets1983,binnigScanningTunneling1987, drakovaTheoreticalModelling2001, hoferTheoriesScanning2003, eversAdvancesChallenges2020, chenIntroductionScanning2021}.
For these studies, an atomically thin insulating layer, \latin{e.g.}, few monolayer sodium chloride NaCl, is often introduced as a buffer between the substrate and the molecule.\cite{reppMoleculesInsulating2005}
This layer helps reduce the influence of the underlying substrate on the molecule.
However, it is usually assumed that the buffer itself has a small influence on STM images, except that the total current is strongly reduced.
Here, we show that the buffer actually strongly influences the wave function of the tunneling electron at the molecule.

STM for molecules can be used for manipulating electrons and photons.
For instance, studies of single free-base phthalocyanine and platinum phthalocyanine (PtPc) molecules have shown upconversion electroluminescence\cite{chenSpinTripletMediatedUpConversion2019, farrukhBiaspolarityDependent2021, GrewalSingleMolecule2022}, where tunneling electrons lead to the emission of photons with larger energy than the applied bias voltage.
Copper phthalocyanine shows negative differential resistivity \cite{tuControllingSingleMolecule2008, siegertEffectsSpin2015, donariniTopographicalFingerprints2012, siegertNonequilibriumSpin2016}, where the total tunneling current is reduced as the bias is increased.
In these cases, the electrons typically tunnel through the molecular electronic transport gap, making tunneling through the gap of particular interest.

Here we focus on how the buffer affects the tunneling through the electronic transport gap.
The purpose is to show that it actually strongly influences any electron tunneling through the transport gap.
This effect can be used to strongly influence the character of the electron wave function at the molecule.
In addition to varying the molecule, the composition and thickness of the buffer then greatly expand the means for manipulating electrons and photons, \latin{e.g.}, in upconversion electroluminescence or negative differential resistivity.
In experiments, typically the buffer is treated as ancillary to a molecule of interest, but we show that it is, in fact, an underestimated and significant means for controlling the tunneling process.

In the spatial range of an adsorbed molecule, the wave function of a tunneling electron can be approximated as a linear combination of molecular orbitals (MOs) \cite{grewalCharacterElectronic2023, grewalScanningTunneling2024}. These MOs couple to the substrate via hopping (overlap) to the buffer states. For a complicated molecule, such as (planar) PtPc treated here, the highest occupied MO (HOMO) and energetically nearby orbitals, like the lowest unoccupied MO (LUMO), tend to have many nodal surfaces because they are orthogonal to many lower-lying MOs. We show that these frontier orbitals then tend to couple inefficiently to the substrate. Low-lying MOs, which have few or even no nodal surface perpendicular to the buffer surface, can have a much more effective coupling to the substrate, which can be orders of magnitude stronger than the HOMO coupling. Although these low-lying MOs are energetically much further away from an electron tunneling through the gap, which strongly reduces their relative importance, they can still play an important role overall in describing the tunneling electron wave function \cite{grewalCharacterElectronic2023, grewalScanningTunneling2024}, even at energies that are only slightly off resonance from the HOMO or LUMO.
The importance of these effects is worth emphasizing because these contributions may be overlooked in theoretical calculations.
It also has a useful experimental consequence.
By varying the buffer composition and thickness, we can then vary the relative importance of different MOs in the tunneling process.

\begin{SCfigure}
\includegraphics[width=240pt]{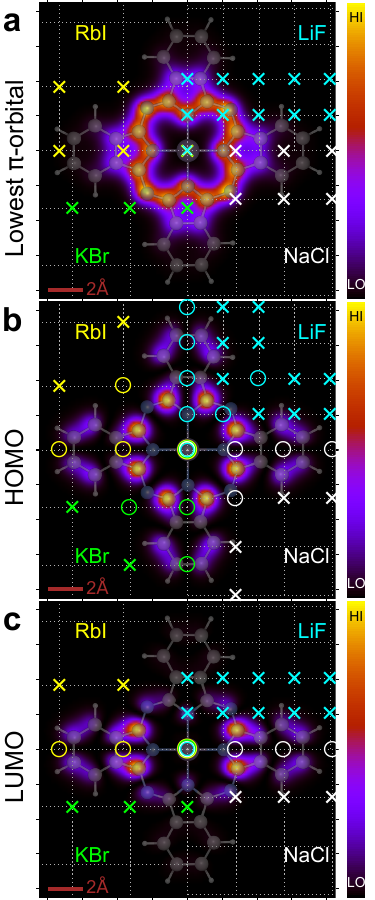}\hspace{1em}
\caption{The norm squared of the PtPc molecule's lowest $\pi$~orbital (top), HOMO (middle) and LUMO (bottom) in a plane between the PtPc molecule and the buffer (at $z=-1$ \AA).
The symbols mark the underlying buffer lattice positions for the LiF (upper right), NaCl (lower right), KBr (lower left) and RbI (upper left) buffers, emphasizing the small number of RbI sites close to the molecule. The Pt atom is assumed to be adsorbed on top of a cation.
The symbol ``o'' marks positions for which the buffer $s$ and $p_z$ orbitals do not couple to the PtPc MO for symmetry reasons, while ``x'' marks positions where coupling is possible.
Because of the many nodal planes of the HOMO, it does not couple to $s$- and $p_z$- orbitals at positions with high symmetry relative to the HOMO.
Due to the large lattice parameter of RbI, sites where the $p_z$ and $s$ orbitals could couple to the HOMO for symmetry reasons are essentially outside the range of the molecule, while for LiF a few such sites are not.}
\label{fig:1}
\end{SCfigure}

Heuristically, the $p_z$ orbitals of the buffer, pointing out of the plane of the buffer towards the molecule, may be expected to play the most important role for the coupling, since the $p_x$ and $p_y$ orbitals point along the plane.
The $s$~orbitals would then be expected to be intermediate between these two extreme cases.
However, the buffer atoms close to the adsorption site of PtPc typically sit at very symmetric positions relative to PtPc, as illustrated in Fig.~\ref{fig:1}. 
From symmetry arguments, it then follows that the $s$ and $p_z$ orbitals on these sites, marked by "o" in the middle panel, cannot couple to the HOMO. 
Counterintuitively, only $s$ and $p_z$ orbitals relatively far from the adsorption site, marked by "x", as well as some $p_x$ and $p_y$ orbitals, couple.
Similar arguments apply also to several other MOs with many nodal surfaces. 
Given these circumstances, what possibilities exist for relatively strong coupling?
Consider the lowest PtPc $\pi$ MO (top panel of Fig.~\ref{fig:1}), which has no nodal surface perpendicular to the buffer surface and therefore couples also to central $s$ and $p_z$ buffer orbitals. This leads to a very efficient coupling to this MO and to other MOs with few nodal surfaces. The buffer, therefore, has a drastic effect on the relative coupling to different MOs by an electron tunneling through the gap and thereby on the character of the tunneling electron. 

The discussion above refers to the coupling of a PtPc MO to a specific orbital on a specific buffer atom. For the coupling to a buffer eigenfunction there are in addition important interference effects between the couplings to different buffer 
atomic orbitals. However, summing over buffer eigenfunctions, these interference effects tend to cancel out. The arguments above therefore address the dominating effects.

This raises a question of how the chemical composition of the buffer influences this coupling.
We consider several different buffers, namely LiF, MgO, NaF, NaCl, KBr, and RbI, all simple salts.
Due to the very different sizes of the ions involved, the lattice parameters span almost a factor of two, from 4.02 \AA \ (LiF), 4.21 \AA \ (MgO), 4.63 \AA \ (NaF), 5.54 \AA \ (NaCl), 6.59 \AA \ (KBr) to 7.34 \AA \ (RbI).
This means that for LiF and MgO, ions in nonsymmetric positions relative to PtPc are closer to the center of the molecule and therefore couple more efficiently to MOs with many nodes than, \latin{e.g.}, RbI.
The choice of buffer, therefore, has a large influence on the character of the wave function of electrons tunneling through the gap and its coupling to different MOs.

An interesting possibility is the tunneling of a substrate electron to an unoccupied MO together with the tunneling of an electron from an occupied MO to the tip in an energy-conserving quantum mechanical process.
The resulting exciton may then decay by emitting a photon.
This process depends crucially on which orbitals form the exciton, as this strongly influences the corresponding dipole matrix element and the coupling to photons.
By varying the buffer, we can influence which orbitals form the exciton and thereby photon emission.
Here, we provide the basis that justifies this anticipation.

We use a tight-binding (TB) formalism, taking into account the substrate (Au, Ag, or Cu), the buffer (LiF, MgO, NaF, NaCl, KBr, or RbI), and the 182 bound orbitals of the PtPc molecule.
The corresponding formalism is described in Supplementary Information (SI).
The manner of electron propagation from the substrate to the molecule is the primary focus of this paper.
The propagation through vacuum is described in cylindrical coordinates and was discussed extensively in Ref.~\citenum{grewalScanningTunneling2024}, as well as in its accompanying SI.
The calculations here follow a similar approach.

We have shown that this TB formalism gives results in very good agreement with experiment. This applies both to the description of the HOMO and LUMO as well as the images in the conduction gap \cite{grewalScanningTunneling2024}. We have also shown that the formalism appropriately describes images for different orientations of the adsorbed molecule, and we have discussed the effects of different adsorption sites \cite{grewalCharacterElectronic2023}. 
\section{Results and discussion}

\subsection{Theoretical formalism}\label{sec:3}

\noindent We study a (planar) PtPc molecule adsorbed on a few layers of a buffer atop a noble metal substrate. 
As earlier \cite{leonAnionicCharacter2022, grewalCharacterElectronic2023, grewalScanningTunneling2024}, we use a TB formalism to describe the substrate-buffer-PtPc molecule complex, essentially following prescriptions of Harrison \cite{harrisonElementaryElectronic1999} (see SI).
We write the corresponding Hamiltonian as
\begin{eqnarray}\label{eq:1}
&&\mathcal{H}=\sum_{i\sigma} \varepsilon_i^{\rm Sub}n_{i\sigma}^{\rm Sub}+\sum_{i\sigma}\varepsilon_i^{\rm Buff}n_{i\sigma}^{\rm Buff}+\sum_{i=1}^{182}\sum_{\sigma} \varepsilon_i^{\rm PtPc}n_{i\sigma}^{\rm PtPc} \nonumber \\
&&+\sum_{ij\sigma}[V_{ij}^{\rm Sub-Buff}(c^{\rm Sub}_{i\sigma})^{\dagger}c^{\rm Buff}_{j\sigma} +\text{h.c.}]  \\
&&+\sum_{ij\sigma}[V_{ij}^{\rm Buff-PtPc}(c^{\rm Buff}_{i\sigma})^{\dagger}c^{\rm PtPc}_{j\sigma}+\text{h.c.}], \nonumber 
\end{eqnarray}
where h.c. is the complex conjugate. The first three terms yield the energies of the substrate states, the buffer states, and the 182 PtPc states, respectively. The last two terms describe the hopping between the substrate and the buffer and between buffer and PtPc, respectively. It has been shown that the conduction band of NaCl is mainly located on the Cl $4s$ level. \cite{clarkAugmentedPlane1968, deboerOriginConduction1999, olssonScanningTunneling2005, leonAnionicCharacter2022}
Here we assume that the conduction band is located on the anion also for the other buffers. 
\section{Results and discussion} 
The values of the various parameters are discussed in SI. We use the Cu(111), Au(111) and Ag(100) substrate surfaces and neglect possible reconstructions. In all cases we assume that PtPc adsorbs on the cation site and with the same orientation as for NaCl on Au (see SI), since in most cases we are not aware of experimental information. This approach focuses on the great importance of the buffer lattice parameter.
For the real system there may be appreciable additional effects due to the specifics of the adsorption for that system, see, e.g., Ref.~\citenum{grewalCharacterElectronic2023}.

This Hamiltonian is used for describing the system inside a matching plane at $z_0=1$ \AA \ outside the the nuclei of the PtPc molecule.
Outside this plane, we introduce cylindrical coordinates with the radial coordinate, $\rho$, the azimuthal angle, $\phi$, and a coordinate perpendicular to the surface, $z$, see SI and Ref.~\citenum{grewalScanningTunneling2024}.
We neglect changes of the potential induced by the tip. We use the theory of Tersoff and Hamann \cite{tersoffTheoryApplication1983, tersoffTheoryScanning1985} to describe the tunneling interaction with the tip.
With these assumptions, the tip does not enter the calculations explicitly. 

Neglecting the attractive potential from the substrate-buffer system for $z>z_0$ as well as the neglect of the potential from the tip tends to overestimate the rapid decay of the wave functions in the vacuum region.
Coulomb interactions are not explicitly included.
The HOMO and LUMO positions are adjusted to the values determined from STM for PtPc on NaCl on Au.
As long as the important PtPc states are the neutral ground-state and states with one extra electron or one hole, Coulomb effects and image potential effects are implicitly included by the use of level positions determined by STM, while, e.g., exciton effects are not included.
For other combinations of buffer and substrate, the HOMO and LUMO alignments may be substantially different, but in most cases, we are not aware of corresponding experimental results.
We have therefore used the same alignments as for NaCl on Au, focusing on the effects of the buffer composition and thickness.

The eigenstates of the Hamiltonian (Eq.~\ref{eq:1}) are calculated.
On the PtPc molecule, an eigenstate $|i\rangle$ of the full system is described as a linear combination of the 182 eigenstates $|\nu\rangle$ of the free PtPc molecule isolated in the gas phase
\begin{equation}\label{eq:1a}
|i\rangle=\sum_{\nu =1}^{182}c^{(i)}_{\nu} |\nu\rangle.
\end{equation}
We can then define a partial density of states on the PtPc molecule
\begin{equation}\label{eq:2}
N_{\nu}(\varepsilon)=\sum_i \left|c^{(i)}_{\nu}\right|^2\delta(\varepsilon-\varepsilon_i)
\end{equation}
and a partial occupancy
\begin{equation}\label{eq:3}
w_{\nu}=C\int_{\varepsilon_1}^{\varepsilon_2} N_{\nu}(\varepsilon){\rm d}\varepsilon.
\end{equation}
The gap of the PtPc model extends from $-1.3$ eV to $1.7$ eV.
We choose $\varepsilon_1=-0.8$ eV and $\varepsilon_2=1.2$ eV, which is purposely a little bit inside the gap.
The prefactor was chosen so that the sum over the 182 weights $w_{\nu}$ add up to unity, that is, 
\begin{equation}\label{eq:4}
    C^{-1}=\sum_{\nu} \int_{\varepsilon_1}^{\varepsilon_2} N_{\nu}(\varepsilon){\rm d}\varepsilon.
\end{equation}

\subsection{Results}\label{sec:4}

We now focus on the MOs built up from $p_z$ orbitals on atoms in the molecule.
The $p_z$ orbitals point towards both the tip and the underlying buffer, and the corresponding MOs are therefore particularly important for the current.
Some of these MOs can be classified according to the number of angular nodal surfaces, $n_a$, and radial nodal surfaces, $n_r$.
Here $n_a$ refers to the number of nodal surfaces which are (approximately) planes through the center of the molecule and perpendicular to the plane of the molecule.
In a similar way, $n_r$ refers to nodal surfaces which are approximately cylinders perpendicular to the plane of the molecule. 
There are also other orbitals that fall outside of this method of classification. 

\begin{SCfigure}
\includegraphics[width=240pt]{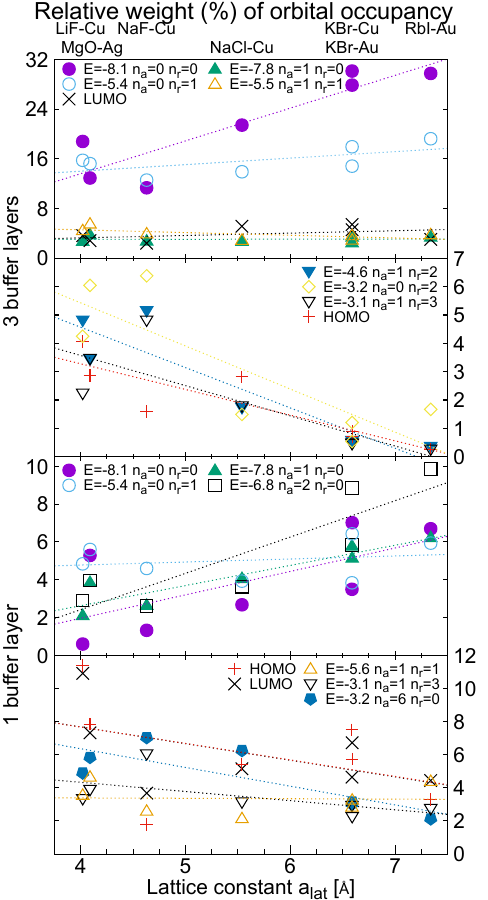}\hspace{1em}
\caption{Partial occupancy $w_{\mu}$ for important PtPc MOs for three (top two panels) and one (bottom two panels) buffer layers integrated over the energy interval $-0.8$ to $1.2$~ eV.
We consider the buffers LiF, MgO, NaF, NaCl, KBr, and RbI on substrates Cu, Ag, or Au, and plot the results as a function of the buffer lattice parameter.
The MOs are labeled by the number of angular nodes, $n_a$, and radial nodes, $n_r$.
Dotted straight lines have been fitted to the results for a given MO.
Observe that the vertical scale is a factor $\approx2.7-4.6$ larger for the top panel than the lower three panels.
The lines are guides for the eye.}
\label{fig:2}
\end{SCfigure}

Fig.~\ref{fig:2} shows the weights of important orbitals for energies, $-0.8\le \varepsilon\le 1.2$ eV, inside the PtPc energy gap for PtPc on a three (top two panels) or one (bottom two panels) layer atomic buffer for different combinations of substrate and buffer.
For degenerate orbitals, we show the weight of one orbital.

We first consider three buffer layers (two top panels). In the top panel the weights of the orbitals increase, stay constant or moderately decrease with the buffer lattice parameter,
while in the second panel the weights rapidly decrease. In the upper panel the number of nodal surfaces is small, with the exception of
the LUMO, discussed later. In particular, the two orbitals with no angular node ($n_a=0$) and with zero ($n_r=0$) or one radial node
($n_r=1$) have large relative weights, which increase with the buffer lattice parameter. In the second panel the weights of the orbitals rapidly drop
with the lattice parameter and become very small for the largest lattice parameters (do note the change in vertical scale from
the top figure). These orbitals typically have more nodal surfaces.

We next consider one monolayer thick buffer (two bottom panels). In the third (fourth) panel we show MOs for which the weights tend to increase (decrease) with the lattice parameter of the buffer.
As for the case of three layers, the weights of MOs with few (many) nodal surfaces tend to increase (decrease) as the lattice parameter increases.
Thus, orbitals with few and many nodal surfaces typically behave in qualitatively different ways for both one and three buffer layers.

In perturbation theory, we might expect the weight of a MO to be given by
\begin{equation}\label{eq:5}
\sum_{-0.8 \le \varepsilon_n\le 1.2} \left|\sum_i\frac{\langle {\rm Subst}  \ n |H| {\rm Buff} \ i \rangle \langle {\rm Buff} \ i|H|\nu\rangle}{(\varepsilon_i -\varepsilon_n) (\varepsilon_{\nu} -\varepsilon_n) }  \right|^2, 
\end{equation}
where $\varepsilon_{\nu}$ is the energy of the PtPc $\nu$th MO, $\varepsilon_i$ is the energy of a buffer state $i$ and $\varepsilon_n$ the 
energy of a substrate state.
The second energy denominator in Eq.~\ref{eq:5} is then in the range $0.5-2.5$ eV for the HOMO or LUMO but about 8 eV for the lowest $\pi$~orbital, \latin{i.e.}, about an order of magnitude larger.
Nevertheless, Fig.~\ref{fig:1} shows that the weight of the lowest $\pi$~orbital is larger.
These results are discussed in the next section.

These results refer to the weights in the plane of the molecule for the different MOs for an electron tunneling through the transport gap.
In Ref.~\citenum{grewalScanningTunneling2024} we focused on the propagation through vacuum outside
the molecule, continuing the MOs into vacuum. It was shown that this propagation, in addition, exponentially favors some components of the wave function 
already strongly favored in Fig.~\ref{fig:1}, \latin{e.g.}, the lowest $\pi$~orbital.

\subsubsection{Effects of the nodal structure of molecular orbitals}\label{sec:5}
\noindent We now discuss why different MOs have very different effective coupling to the substrate.
Fig.~\ref{fig:1} shows the lowest $\pi$~orbital, the HOMO, and one of the two degenerate LUMOs in a plane between the PtPc molecule and the buffer.
In addition, the figure shows the sites of the LiF (upper right), NaCl (lower right), KBr (lower left) and RbI (upper left corner) buffer layer.
Based on these figures, we now discuss the strength of the coupling of the three MOs to different buffers.
One may expect the coupling of PtPc to the buffer $s$ and $p_z$ orbitals to dominate over the coupling to the $p_x$ and $p_y$ orbitals, which point along the buffer surface.
We, therefore, focus on the former set of orbitals. 

Let us first consider the lowest $\pi$~orbital, which has no nodes except in the plane of the molecule.
This orbital then couples to all $s$ and $p_z$ orbitals of the buffer within the spatial extension of the molecule.
Because of the larger lattice parameter of, \latin{e.g.}, RbI than, \latin{e.g.}, LiF, there is coupling to fewer buffer orbitals for RbI, even though there is still a substantial coupling also for RbI.

We next consider the HOMO.
Because of the many nodal planes perpendicular to the buffer, the HOMO does not couple to the $s$ and $p_z$ orbitals on many sites in symmetric positions relative to PtPc.
These sites are marked by ``0'' (see Fig. \ref{fig:1}).
In particular, this zero coupling applies to all positions close to the center.
Orbitals on sites further out are in non-symmetric positions and can couple.
However, only a few of these orbitals are spatially within the extension of the molecule, so they mainly couple to the outer parts of the HOMO.
This specific manner of coupling is particularly important for buffers with a large lattice parameter.
Fig.~\ref{fig:1} illustrates that for LiF there are still a few such buffer sites inside the PtPc molecule, whereas for RbI these sites are really only in contact with the outer edges of the PtPc.
Due to the efficient suppression of the coupling to buffer $s$ and $p_z$ orbitals, the coupling to $p_x$ and $p_y$ orbitals becomes relatively more important.

Each of the LUMO orbitals has a node along one of the coordinate axis, and therefore does not couple to $s$ or $p_z$ orbitals on that axes.
The sites for which $s$ and $p_z$ orbitals can couple are then rather far out, particularly for RbI.
These sites have weak couplings to the LUMOs, in a similar manner to the HOMO. 

To illustrate these effects, we first focus on the hopping matrix elements between a MO $\nu$ and a buffer orbital $i$.
The sum of the squares of these matrix elements is then calculated as
\begin{equation}\label{eq:6}
A_{\nu}=\sum_i |\langle {\rm  \ Buff}~ i|H|\nu\rangle|^2.
\end{equation}

\begin{SCfigure}
\includegraphics[width=240 pt]{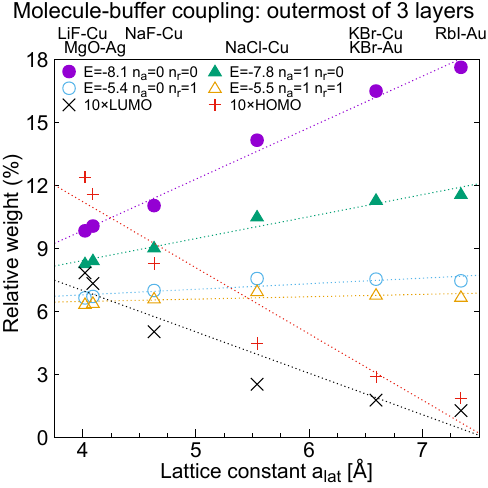}\hspace{1em}
\caption{Coupling, $A_{\nu}$,  of MOs to outermost buffer layer (Eq.~\ref{eq:6}) for different buffers as a function of the buffer lattice parameter $a_{\rm lat}$ and for the case of three buffer layers.
Results are shown for a few low-lying MOs with few nodal surfaces and strong coupling, and to the HOMO and LUMO.
Observe that the coupling strengths to the HOMO and LUMO are multiplied by a factor 10, illustrating the weak coupling to these orbitals.
For the two-fold degenerate orbitals we show the coupling to one of the orbitals.
The dotted straight lines are guides for the eye.}
\label{fig:3}
\end{SCfigure}

The results are shown in Fig.~\ref{fig:3} for different buffers.
The figure illustrates how the couplings to the HOMO and LUMO are strongly reduced relative to the coupling to several low-lying $\pi$~orbitals, especially as the lattice parameter of the buffer, $a_{\rm lat}$, is increased. In Fig.~\ref{fig:1} the LUMO nevertheless only shows a weak lattice parameter dependence, due to compensating effects in the full treatment, beyond perturbation theory, of the buffer propagation and the buffer-substrate coupling.

We next focus on the propagation through the buffer and introduce  
\begin{equation}\label{eq:7}
B_{\nu}=\sum_{\gamma \subset {\rm layer} \ 1} \left|\sum_i \frac{\left\langle \gamma|{\rm Buff} \ i\right\rangle \langle {\rm Buff} \ i|H|\nu\rangle }{ \varepsilon_i}\right|^2.
\end{equation}

\begin{SCfigure}
\includegraphics[width=240 pt]{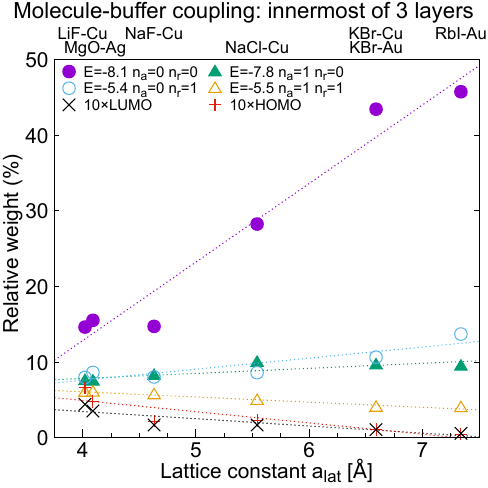}\hspace{1em}
\caption{Coupling, $B_{\nu}$, of MOs to innermost buffer layer (Eq.~\ref{eq:7}) for different buffers as a function of the buffer lattice parameter $a_{\rm 
lat}$ and for three buffer layers.
Observe that the coupling strengths to the HOMO and LUMO are multiplied by a factor 10. For the two-fold degenerate orbitals we show the coupling to one of the orbitals.
The dotted straight lines are guides for the eye.}
\label{fig:4}
\end{SCfigure}

This expression is guided by the perturbation theory in Eq.~\ref{eq:5} and related to the coupling of the $\nu$th MO to the buffer layer closest to the substrate (layer 1).
The energy denominator in Eq.~\ref{eq:5} containing the MO energy is not yet taken into account.
In the denominator, including the buffer energy, $\varepsilon_i$, we have neglected the substrate energy, $\varepsilon_n$, since $|\varepsilon_n| \ll|\varepsilon_i|$ in Eq.~\ref{eq:5}.
The results are shown in Fig.~\ref{fig:4}.
Comparison with Fig.~\ref{fig:3} shows that the propagation through the buffer, in particular, favors the coupling to the lowest $\pi$~orbital.
This explains why the coupling to the lowest $\pi$~orbital dominates for three buffer layers, while the couplings to several low-lying orbitals are comparable for one buffer layer.
While this perturbation theory expression for $B$ may lack quantitative accuracy, it effectively illustrates qualitative factors that render the contributions from low-lying orbitals crucial, despite their very unfavorable energy denominators.

\subsection{Effects of vacuum propagation for tunneling through the transport gap}\label{sec:6}

\noindent Up to this point, we have discussed tunneling through the transport gap, focusing on the different components of the wave function at the adsorbed molecule.
We now shift our focus to the image seen when the tip is scanned over the molecule.

\begin{SCfigure}
\includegraphics[width=240 pt]{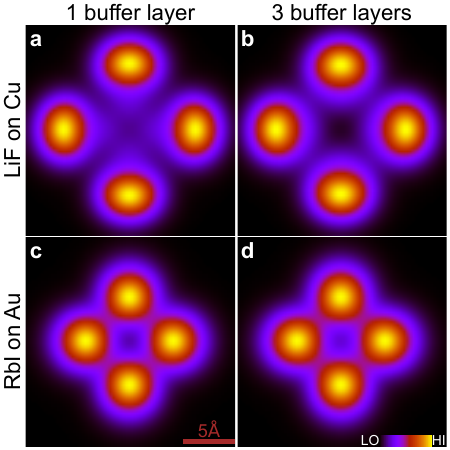}\hspace{1em}
\caption{Images at $z-z_0=6$ \AA \ and $\varepsilon=0$ for PtPc on LiF on Cu (top row) and on RbI on Au (second row) for a buffer with one layer (left) or three layers (right). All panels show an area of $ 20 \times 20~ \text{\AA}^2$.}
\label{fig:5}
\end{SCfigure}

Fig.~\ref{fig:5} shows results for the two buffers with the smallest and largest lattice parameter (LiF and RbI) and for two thicknesses, one or three layers.
It is striking that the differences are rather small.
A pattern with a somewhat wider spacing is found for LiF, although it has a smaller lattice parameter.
It is interesting that the spatial range of the image does not follow the lattice parameter of the buffer but, rather, is determined by the specific coupling to the substrate \latin{via} the buffer.

We demonstrated earlier that the weight of some components at the molecule depends very strongly on the buffer.
For instance, the weight of the HOMO varies by one order of magnitude between the different buffers.
However, the weight of components with a more complicated nodal structure, such as the HOMO, have rather small weights already at the adsorbed molecule.
These components are further strongly (exponentially) damped during propagation through vacuum relative to, \latin{e.g.}, the lowest $\pi$~orbital \cite{grewalScanningTunneling2024}. 

The image at the tip of an electron tunneling through the transport gap is, therefore, dominated by components with very simple nodal structures \cite{grewalCharacterElectronic2023,grewalScanningTunneling2024}, primarily components with no or one angular node.
The components with more complicated angular structures play just a small role.
The strong variation of the weights of these more complicated orbitals with the buffer lattice parameter then plays an overall small role for the image at the tip.

The small difference between images for different buffers might suggest the traditional picture, where the buffer just isolates the molecule from the substrate but otherwise does not influence the image much.
However, this is a common misinterpretation.
Rather, as discussed above, all buffers actually strongly suppresses components with a complicated nodal structure.
The image in the transport gap then looks similar for different buffers exactly because only a few components with 
simple nodal structures dominate the image. However, as discussed above, in the plane of the molecule, where manipulation 
of electrons and photons take place, more complicated components can play a significant role.

\section{Conclusions}\label{sec:7}

\noindent We have studied the tunneling through the transport gap of a molecule, focusing on effects of a buffer between the molecule and the substrate.
We focused on the wave function of the tunneling electron in the spatial region of the molecule, and expressed it as a linear combination of MOs of the molecule.

We demonstrated that the transport through the buffer greatly favors $\pi$-MOs with a simple nodal structure compared with the HOMO, LUMO, and other MOs with more complex nodal structures.
Despite the HOMO and LUMO being hugely favored by their proximity in energy to the tunneling electron, MOs at substantially lower energies tend to dominate the wave function of an electron tunneling through the transport gap.

We found that the lattice parameter of the buffer plays a crucial role, and buffers with large lattice parameters strongly disfavor MOs with a complex nodal structure.
For instance, the relative weight of the HOMO in the transport gap is suppressed by one order of magnitude in going from a LiF to a RbI buffer.

Additionally, we found that buffers with just one atomic layer influence the electrons in different ways than buffers with more layers. 

During the propagation through vacuum, the components of the wave function corresponding to a few low-lying $\pi$~orbitals, already
favored by the buffer, are further exponentially favored by the vacuum propagation. Just a couple of components then dominate
the image so much that it becomes relatively independent of the buffer (see Fig.~\ref{fig:5}).   

For applications involving manipulation of electrons, such as photon emission, upconversion, and negative differential 
resistivity, however, the electron wave function in the spatial region of the molecule plays a crucial role.
In this region of space additional components of the wave function can also play an appreciable role by appropriate choice
of buffer (see Fig.~\ref{fig:3}).
Traditionally, these applications relied on finding an appropriate choice of an adsorbed molecule.
The results above, however, provide two additional possibilities for influencing these processes --- the choice of the buffer chemical composition, particularly the buffer lattice parameter, and the number of buffer atomic layers.

\begin{suppinfo}
The online Supporting Information is available free of charge.
\begin{itemize}
    \item Tight-binding model, Substrates, Buffers, Model of PtPc, Vacuum propagation
\end{itemize}
\end{suppinfo}

\section{Data availability}
The data supporting this study's findings are available from the corresponding authors upon reasonable request.

\section{Competing interests statement}
The authors declare no competing interests.

\section{Author contributions statement}
All authors participated in the analysis and discussion of the results, and in the writing of the manuscript. O.G. conceived and designed the research program and performed the calculations.

\section{Funding}
Open access funded by Max-Planck Society.

\section{Acknowledgment}
We would like to thank K. Kuhnke and G. Li Manni for helpful comments and discussions.

\providecommand{\latin}[1]{#1}
\makeatletter
\providecommand{\doi}
  {\begingroup\let\do\@makeother\dospecials
  \catcode`\{=1 \catcode`\}=2 \doi@aux}
\providecommand{\doi@aux}[1]{\endgroup\texttt{#1}}
\makeatother
\providecommand*\mcitethebibliography{\thebibliography}
\csname @ifundefined\endcsname{endmcitethebibliography}
  {\let\endmcitethebibliography\endthebibliography}{}

\clearpage
\begin{center}
    \textbf{\Large Supplementary Information}
\end{center}

\noindent We describe the tight-binding models for PtPc on different buffers on a Cu, Ag, or Au substrate, generalizing earlier descriptions for NaCl on Au \cite{grewalCharacterElectronic2023A, grewalScanningTunneling2024A} to the new substrates and buffers.
We generally use the tight-binding formalism of Harrison \cite{harrisonElementaryElectronic1999A} to obtain nearest-neighbor hopping integrals between basis functions located on the atoms.

The buffer and substrate are incommensurate in all cases except MgO on Ag.
While the nearest neighbors are well-defined for hopping within the substrate, the buffer, and PtPc, they are ill-defined for hopping between the substrate and the buffer and between the buffer and PtPc.
We, therefore, use a smooth distance-dependent cutoff of the Harrison prescription for these two latter cases.
The Harrison hopping integrals are multiplied by a factor 
\begin{equation}\label{eq:s1}
e^{-(d-d_i)^2/\lambda_i^2}, \hskip0.2cm i= {\rm B \ or \  M},
\end{equation}
for these hoppings.

For the substrate-buffer hopping B, $d$ is the distance between a substrate atom and a buffer atom at the substrate-buffer interface, and $d_B$ is the separation between the adjacent substrate-buffer planes.
The parameter $\lambda_B$ is chosen so that, on average for each buffer atom, the factors in Eq.~\ref{eq:s1} add up to four, representing a typical number of nearest neighbors in a neighboring plane.
For the buffer-PtPc hopping, $d$ is the distance between a PtPc atom and a buffer atom in the outermost plane, and $d_M$ is the separation between the outermost buffer plane and the plane of PtPc.
Again, $\lambda_M$ is chosen so that, on average, these factors add up to four for each PtPc atom.

The orbital energies are essentially obtained from Harrison \cite{harrisonElementaryElectronic1999A}.
However, we have modified some energies to correct known inaccuracies.
Thus, we shift the energy of the substrate $d$-orbital so that the distance of the top of the $d$-band to the Fermi energy agrees with results in the literature.
Finally, we shift all states uniformly so that the Fermi energy $E_F=0$.

We have shown that for NaCl the conduction band primarily has anion $s$ character \cite{leonAnionicCharacter2022A}.
We assume that the corresponding results also apply to MgO and the other alkali halides.
Thus, we adjust the level energies of the buffers so that the conduction band has primarily anion $s$, and the experimental gaps are reproduced.
These level energies then also include the effect of the Madelung potential, while image effects due to the substrate were neglected.
Finally, we shift all levels uniformly, to obtain the correct location of the top of the valence band relative to $E_F$, when this location is known.

The parameters of PtPc are discussed below.
The tight-binding wave functions are then matched to the exponentially decaying wave function in the vacuum region outside PtPc as described in Ref.~\citenum{grewalScanningTunneling2024A}.

Whenever experimental results are not known for some parameter, we use the parameters for NaCl on Au as a guide.
This then puts the focus on known differences between the different systems, in particular, the different lattice parameters of the buffers.

Although we use models for the substrate with a large number of atoms, the detailed results depend on the precise arrangement of the atoms in the substrate.
In most cases we then perform calculations with different number of layers and different number of atoms per layer, giving a total number of substrate atoms of around $3000-4500$.
For the buffer we typically use 324 atoms per layer.
This then leads to matrices of sizes up to about $45000 \times 45000$.
The results of these different calculations are then averaged.

\section*{Substrates}
\noindent We use the lattice parameters $a_{\rm Cu}=3.60$ \AA \ \cite{daveyPrecisionMeasurements1925}, $a_{\rm Ag}=4.09$ \AA \ \cite{chenPropertiesTwodimensional2014} and $a_{\rm Au}=4.07$ \AA \ \cite{daveyPrecisionMeasurements1925}.
The top of the $d$-band is put at 2.0 eV,\cite{rothAngleResolved2016} 4.0 eV \cite{rothAngleResolved2018} and 1.6 eV \cite{sheverdyaevaEnergymomentumMapping2016} below the Fermi energy for Cu, Ag, and Au, respectively.
We use the Cu(111), Ag(100), and Au(111) surfaces.
The $s$ and $d$ level energies were obtained from Harrison, and we added a $p$ level about 5 eV above the $s$ level.
Finally, the Harrison parameters were slightly modified to obtain the desired $d$-band position relative to $E_F$.

\section*{Buffers}
\subsection*{LiF}
\noindent We use the lattice parameter is 4.02 \AA \ \cite{sirdeshmukhStructurerelatedParameters2001}.
Starting from the known separation of the innermost KBr layer on Cu, 2.90 \AA \ \cite{schulzendorfAlteringProperties2019}, we estimate the corresponding distance for LiF on Cu by using Shannon's ionic radii \cite{shannonRevisedEffective1976}.
Assuming that the larger F$^{-}$ ions determine the separation, we obtain the estimate $2.90+1.33-1.96=2.27$ \AA.
The band gap of LiF is 13.6 eV \cite{roesslerElectronicSpectrum1967}, and the Li and F parameters were adjusted accordingly.
Based on Ref.~\citenum{watkinsVacuumLevel2003} we estimate that the top of the valence band of a thin film of LiF on Au is about 7 eV below the Au Fermi energy and use the same alignment for LiF on Cu. 

\subsection*{MgO}
\noindent The MgO lattice parameter is adjusted from 4.21 \AA \ to the value for Ag \cite{chenPropertiesTwodimensional2014}, $a_{\rm MgO}=a_{\rm Ag}=4.09$ \AA, to make the two lattices commensurate.
We use the Ag-MgO separation $d_{\rm Ag-MgO}= 2.62$  \AA \ \cite{chenPropertiesTwodimensional2014}.
O atoms are  put on top of Ag atoms \cite{schintkeInsulatorsUltrathin2004}.
The Mg atoms then sit in hollow positions.
After these adjustments, the density of O atoms and Ag atoms per layer is the same.
The O and Mg levels are adjusted so that the bulk band gap of MgO is 7.8 eV \cite{heoBandGap2015}.
All MgO levels are shifted so that the top of the O $2p$ band is 4 eV below $E_F$ \cite{schintkeInsulatorsUltrathin2004}.

\subsection*{NaF}
\noindent We use the lattice parameter 4.63 \AA \ \cite{sirdeshmukhStructurerelatedParameters2001}.
The band gap of NaF is 11.5 eV \cite{wasada-tsutsuiElectronicBand2002}.
Based on Shannon's ionic radii \cite{shannonRevisedEffective1976}, we reduce the distance of KBr on Cu $2.90$ \AA \ to $2.90+1.33-1.96=2.29$ \AA.
The band gap of NaF is 11.5 eV \cite{wasada-tsutsuiElectronicBand2002}.
We put the top of the valence band at $-6$ eV, so that the substrate Fermi energy is approximately located in the middle of the gap.

\subsection*{NaCl}
\noindent Calculations find that for a three layer NaCl film on Au the lattice parameter is reduced from the NaCl bulk value to 5.54 \AA \ \cite{chenPropertiesTwodimensional2014}, the value used here. 
We use the calculated separation $d_{\rm Au-NaCl}=3.12$ \AA \ between the Au surface and the NaCl film \cite{chenPropertiesTwodimensional2014}.
The band gap of NaCl is 8.5 eV \cite{pooleElectronicBand1975}.
Based on GW \cite{hedinNewMethod1965} calculations \cite{wangQuantumDots2017}, we put the top of the NaCl valence band 5 eV below the $E_F$.

\subsection*{KBr}
\noindent For KBr we use the lattice parameter $a_{\rm KBr}=6.60$ \AA \ \cite{KBrCrystal}. 
We use  the calculated separation $d_{\rm Cu-KBr}=2.90$ \AA \ \cite{schulzendorfAlteringProperties2019} between the KBr film and the Cu substrate.
For KBr on Au we use the separation for KBr on Cu and add the difference in Cu and Au metallic radii \cite{wellsStructuralInorganic1984} to obtain the separation $d_{\rm Au-KBr}=3.06$ \AA.
The KBr parameters are adjusted so that the band gap is 7.6 eV \cite{arvesonThermallyInduced2018}.
Since we do not know the position of the KBr valence band with respect to $E_F$, we put the top of the band at 5 eV below $E_F$, as for NaCl on Au.

\subsection*{RbI}
\noindent For RbI we use the lattice parameter 7.34 \AA \ \cite{RbICrystal}.
Based on the separation of NaCl from Au and the differences in ionic radii between Cl and I \cite{shannonRevisedEffective1976} we use the separation 3.51 \AA \ between the RbI and Au planes.
The band gap of RbI is 6.1 eV \cite{RubidiumIodide}.
For lack of experimental results, we put the top of the valence band at 4 eV below $E_F$.
This puts the top somewhat higher than for NaCl on Au.
Due to the smaller gap for RbI we used this value to avoid that the gap is too asymmetric relative to the Fermi energy.

\section*{Model of PtPc}
\noindent We study the absorbed PtPc molecule.
The coordinates of PtPc are obtained from a density functional calculation. 
The tight-binding parameters are obtained from Harrison \cite{harrisonElementaryElectronic1999A} and are given in Table~\ref{table:1}.
For the H atoms we include the $1s$ level at the energy $-13.6$ eV (not given by Harrison).
We then slightly shifted the HOMO and LUMO for each substrate-buffer-PtPc combination so that the PtPc HOMO forms a resonance at $-1.3$~ eV and the LUMO a resonance at 1.7 eV, in agreement with experiment for PtPc on NaCl on Au.
Since we are not aware of experimental results for HOMO and LUMO positions in most of the other combinations of substrate and buffer, this alignment was used in all cases.
Adjusting the HOMO and LUMO energies to experimentally observed positions means that various effects are implicitly included, e.g.,
charge rearrangement in PtPc in the presence of a LUMO electron or a HOMO hole, as well as polarization of the buffer due to charging of the molecule.

For PtPc on NaCl on Au \cite{grewalCharacterElectronic2023A, grewalScanningTunneling2024A}, we follow Miwa {\it et al.} 
\cite{miwaEffectsMoleculeinsulator2016} and use the separation 3.4 \AA \ between the the molecule and the NaCl film, absorbed on top of a Na atom.
The four arms of PtPc are along the NaCl (100) directions.
For the hopping between the molecule and the buffer we use an exponential cut-off as described above so that, on average, each atom in PtPc hops to four atoms in the buffer.
The Au slab breaks the four-fold symmetry of PtPc which has been reintroduced in the results.

\renewcommand{\arraystretch}{1.34}
\begin{table}[t]
\caption{Level energies used in the models. The top three lines give substrates parameters, the following 12 lines  buffer parameters and the last four lines PtPc parameters.}
\label{table:1}
\begin{tabular}{lccc}
\hline
\hline
Element & $s$ & $p$ & $d$ \\
\hline
Cu (4s, 4p, 3d)  &   6.0   &  11.0  & $-5.7$   \\
Ag (5s, 5p, 4d)  &   4.1   &   9.0  & $-5.9$  \\
Au (6s, 6p, 5d)  &   4.1   &   9.1  & $-3.7$   \\
Li (2s,2p)       &  23.0   &   29.0 & $-$    \\
Na (NaF)(3s, 3p) &  18.1   &  22.1  & $-$    \\
Na (NaCl)(3s, 3p)&  12.8   &  16.8  & $-$      \\
Mg (3s, 3p)      &  19.5   &  23.8  & $-$      \\
K  (4s, 4p)      &  10.0   &  14.0  & $-$      \\
Rb (5s, 5p)      &   8.5   &  12.5  & $-$      \\
O  (3s, 2p)      &  17.1   &  $-4.0$  & $-$      \\
F (LiF)(3s, 2p)  &  20.2   &  $-7.0$  & $-$      \\
F (NaF)(3s, 2p)  &  15.5   &  $-6.0$  & $-$      \\
Cl (4s, 3p)      &  10.2   &  $-5.0$  & $-$      \\
Br (5s, 4p)      &   6.8   &  $-5.0$  & $-$      \\
I  (6s, 5p)      &   5.3   &  $-4.0$  & $-$      \\
C  (2s, 2p)      &$-19.38$   & $-11.07$ & $-$      \\
N  (2s, 2p)      &$-26.22$   &  13.84 & $-$      \\
Pt (6s, 5d)      & $-6.85$   & $-$      & $-16.47$ \\
H  (1s)          & $-13.61$  & $-$      & $-$      \\
\hline
\hline \\
\end{tabular}
\end{table}

We are not aware of experiments for PtPc giving the separation to the buffer or the orientation of PtPc in most other cases.
We then construct models which follow PtPc on NaCl on Au as much as possible.
In particular, we assume that PtPc is absorbed on the positive ion and has the same orientation as for PtPc on NaCl on Au.
However, based on Shannon's ``ionic radii'',\cite{shannonRevisedEffective1976} we modify the separation to the buffer according to the size of the anion in the buffer. 

The difference in the results between the different cases will then be primarily due to the choice of buffer and substrate.
Should later experiments show that PtPc is absorbed on a different site or has a different orientation, our calculations in Ref.~\citenum{grewalCharacterElectronic2023A} provide guidance about the corresponding changes.

\section*{Vacuum propagation}
\noindent The tight-binding description above is used outwards to a distance $z_0=1$ \AA \ above the molecule.
For the description outside this plane we follow Ref.~\citenum{grewalScanningTunneling2024A}.
For $z>z_0$, we assume a constant potential, $V_0$, inside a cylinder with radius $\rho_0=12$ \AA, and infinite outside.
We use the work function $V_0=4.3$ eV \cite{liSpontaneousDoping2015} and the substrate Fermi energy is used as energy zero.
Then
\begin{equation}\label{eq:2.1}
V(\rho,\phi,z)=\begin{cases}V_0,& \text{if}~\rho \le \rho_0~\text{and}~z\ge z_0;\\
\infty & \text{if}~\rho>\rho_0~\text{and}~z\ge z_0.
\end{cases}
\end{equation}
The cylinder radius (12 \AA) is much larger than the distance from the cylinder axis to the outermost H atoms (7.6 \AA).
We introduce the Schr\"odinger equation for an energy $E(<V_0)$

\begin{eqnarray}\label{eq:2.2}
\left[-\left({\frac{\partial^2}{\partial z^2}}+{\frac{1}{\rho}}{\frac{\partial}{\partial \rho}}+{\frac{\partial^2}{\partial\rho^2}}\right)+ {\frac{1}{\rho^2}}{\frac{\partial ^2}{\partial \phi^2}}+V(\rho,\phi,z)\right]\psi(\rho,\phi,z) = E\psi(\rho,\phi,z)
\end{eqnarray}
The solution is given by   
\begin{eqnarray}\label{eq:2.2a}
\psi(\rho,\phi,z)=\sum_{mi}\left[c_{mi}^{\rm (s)}{\rm sin}(m \phi) +c_{mi}^{\rm (c)} {\rm cos}(m\phi)\right]J_m\left[k_{mi}\rho\right]e^{-\kappa_{mi} z},
\end{eqnarray}

where $m(\ge 0)$ is an integer and $J_m$ is an integer Bessel function.
The coefficients $k_{mi}$ are defined so that $J_m[k_{mi}\rho_0]=0$, to guarantee that the wave function is zero for $\rho=\rho_0$.
To obtain the correct energy, $E$, we require
\begin{equation}\label{eq:2.2b}
[\kappa_{mi}]^2=[k_{mi}]^2-(E-V_0).
\end{equation}
All energies are expressed in Ryd=13.6 eV and lengths in Bohr radii $a_0=0.529$ \AA.
The functions sin$(m\phi)$, $m\ge 1$ and cos$(m\phi)$, $m\ge 0$  describe the dependence on the azimuthal angle $\phi$.
For a given value of $m$, the Bessel functions $J_m[k_{mi}\rho]$, $i=1, 2, \dots$ describe the radial behavior.
Finally, exp$(-\kappa_{mi} z)$ gives the exponential decay in the $z$-direction.
These vacuum solutions are matched continuously to the tight-binding solutions for substrate-barrier-molecule complex.

\clearpage
\input{references_si.bbl}

\end{document}

%% file: references_si.bbl
\providecommand{\latin}[1]{#1}
\makeatletter
\providecommand{\doi}
  {\begingroup\let\do\@makeother\dospecials
  \catcode`\{=1 \catcode`\}=2 \doi@aux}
\providecommand{\doi@aux}[1]{\endgroup\texttt{#1}}
\makeatother
\providecommand*\mcitethebibliography{\thebibliography}
\csname @ifundefined\endcsname{endmcitethebibliography}
  {\let\endmcitethebibliography\endthebibliography}{}